# Design of a High-Resolution Multifocal LIDAR: Enabling higher resolution beyond the laser pulse rise time

Koray Ürkmen and Emre Yüce

*Abstract*—3D imaging is increasingly impacting areas such as space, defense, automation, medical and automotive industries. The most well-known optical 3D imaging systems are LIDAR systems that rely on Time of Flight (ToF) measurement. The depth resolution of the LIDAR systems is limited by the rise time of the illumination signal, bandwidth of the detector systems, and the signal-to-noise ratio. The focal gradients can be utilized to obtain 3D images from 2D camera captures. In this study, we combine the basic LIDAR techniques with a physical measurement of focal gradient changes that are caused by the depth of field, to obtain a high-resolution 3D image using spatially separated single-pixel detectors. The system that we introduce here enables depth resolution that goes beyond the rise time of the incident laser pulse.

## I. INTRODUCTION

THE basic working principle of 3D LIDAR systems is based on illuminating the target with a known signal shape and measuring the time of the returning signal to the detector. The speed of light is almost constant in air, and the time difference between the outgoing and return pulse can be used for measuring the distance between the target and LIDAR. The depth resolution is determined by the Time of Flight (ToF) measurement precision which is limited by the rise time of the signal pulse and the bandwidth of the detector systems. To achieve centimeter depth resolution by the traditional LIDAR systems, pulse rise times in picoseconds, as well as larger detector bandwidths in GHz, are required [1]. Moreover, at an increased distance the signal-to-noise ratio (SNR) decreases. These further limit the depth resolution. There are also studies for increasing the depth resolution of the LIDAR systems without decreasing signal pulse rise time or increasing detector bandwidths where they still need signal rise time times in nanoseconds and bandwidths in of the order of GHz to achieve mm-scale resolutions [6]. Also, modulation techniques can be used for higher depth resolution but the maximum measurement range is limited by the ambiguity in these applications [9, 12].

The depth data can also be derived by de-focusing the image on the image plane [4] or by using Shape-from Techniques such as deriving the depth information from the texture, shading, contours etc. [11]. It is known that the image of closer targets is reconstructed at a longer image plane. Based on the change in target distance the distance between the image plane and the lens will change and this causes de-focusing or blurring of the image. There exist several algorithms for estimating the depth data on the target plane from the amount of the blurring effect on the image plane [2,3]. The blurring effect also depends on the circle of confusion size, which can be resolved by the detector on the image plane, and respectively depth of field.

However, these algorithms mainly require a 2D camera that does not provide an avalanche process at a low cost which is essential for the extensive deployment of LIDARs in various fields.

In this work, we combine the LIDAR system with the separate imaging capability at the different image plane distances to increase the depth resolution without complex laser sources or high bandwidth Focal Plane Array (FPA) systems. The fundamentals of our approach are based on using physically separated single-pixel detectors at the image plane and have been shown to obtain multifocal images in confocal microscopy systems [5]. We show that high depth resolution LIDAR can be developed by using multiple detectors which are positioned at different positions in the image plane. Our detection scheme provides the means to obtain increased resolution in-depth at a reduced cost, see patent application [8].

## II. THEORETICAL BACKGROUND

In our design, the well-known LIDAR ToF technique is combined with the separate imaging capability at the different image plane distances. Here, we aim to measure reference point distance on the target with ToF measurement than getting the relative depth information from the constructed images at the different image planes. After the first measurement that is performed via TOF, the depth resolution is then increased by the number of image planes for each of which a detector is required. The resolution of the reference point distance on the target can be calculated by the simple ToF equation which is;

$$d = \frac{c\Delta t}{2} \qquad (2.1)$$

Emre Yüce and Koray Ürkmen are with Programmable Photonics Group, Department of Physics, Middle East Technical University, 06800 Ankara, Turkey and Micro and Nanotechnology Program, Middle East Technical University, 06800 Ankara, Turkey. (e-mail: eyuce@metu.edu.tr, kurkmen@gmail.com ).

where d is the distance of the reference point on the image, c is the speed of light and Δt the time difference between the ignition of the signal and receiving it back at the LIDAR detector. It is also known that the resolution (σr) of a TOF system depends on , trise is the rise time of the light pulse and SNR is the signal-to-noise ratio, as given in equation (2.2) [7];

$$\sigma_r \propto \frac{c}{2} \cdot t_{rise} \cdot \frac{1}{\sqrt{SNR}} \qquad (2.2)$$

In equation. 2.3 the depth resolution is limited by the rise time of the pulse (trise). On the other hand, this is valid for a fixed image plane. The relative depth information can be obtained by different detectors located on different image planes. In this method, the depth resolution is limited by the depth of field which is illustrated in Fig.1. Here, f is the effective focal length of the lens system; D is target distance; C is circle of confusion diameter; N is f# of the lens system, d1 is the nearest object distance from the D, d2 is the farthest object distance from the D and so d is the depth of field. To get a higher depth resolution, it is aimed to get a smaller depth of field which enables distinguishing images that correspond to different target depths. For this purpose, it is required to increase the effective focal length of the optical system which can be achieved via choosing a small f# and measurement range.

$$d1 + d2 = d = \frac{2NCD^2}{f^2} \qquad (2.3)$$

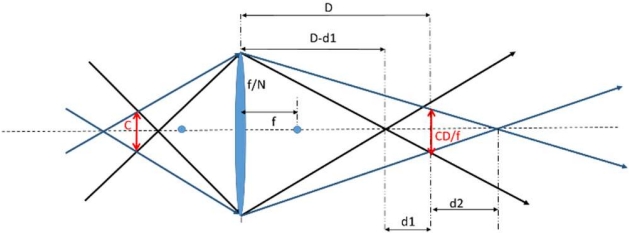

**Fig. 1.** The schematic representation of the depth of field.

### III. INCREASING DEPTH RESOLUTION:

To increase the depth, an FPA sensor is used for image construction. To further increase the depth resolution, the low field of depth is aimed that increases the blurring effect at different depths on the target. A high-resolution positioner is used for changing the image plane location on the optical axis. A piezoelectric stage can be used for this purpose for the realization of the designed scheme. In this study, not only the peak point of the encircled energy (intensity) on each pixel, but also the intensity changes on each pixel via different image plane locations are used for estimating the depth data more precisely [8].

In our first scheme (Fig.2), the light source is used for illumination of the target in pulsed mode for the first ToF measurement and then the higher resolution depth measurement is obtained via focal shifting. The detector is used for the first ToF measurement to find the reference target distance. Beam Splitter is used to steer the rays to the detector plane for first ToF measurement then steering the rays to FPA (image plane) for depth measurement from focal shifting. High-resolution positioner system is used for changing the position of the image plane on the optical axis to create a focusing or defocusing effect.

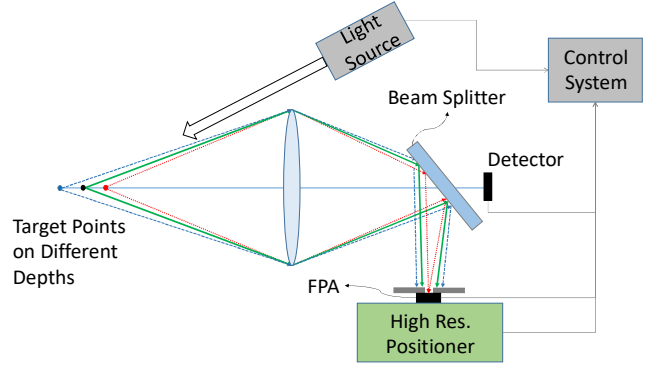

**Fig. 2.** Schematic diagram of the design principle for the increased spatial and depth resolution method. The design scheme relies on moving the FPA sensor.

For the increased spatial and depth resolution method, a Double-Gauss imaging optic has been designed and used for the model of the encircled energy changes due to object depth on the target plane. The parameters of the optical system are given in TABLE I below.

TABLE I
INITIAL DESIGN PARAMETERS OF OPTICAL SYSTEM

| Parameter | Design Data |
|---|---|
| Effective Focal Length (f) | 167 mm |
| Target Distance | 8 m |
| Clear Aperture | 40 mm |
| Pixel Pitch | 4,3 μm |
| FPA Spatial Resolution | 1280 x 960 |

$$d = \frac{2.(^{167}/_{40}).0{,}0043.(8000)^2}{167^2} \qquad (3.1)$$
$$= 82{,}397 \text{ mm}$$

$$the\ IFOV = \tan^{-1}\frac{Pixel\ Pitch}{2*f} \qquad (3.2)$$
$$= \tan^{-1}\frac{0{,}0043}{2*167} \approx 0{,}0015°$$

$$FOV_{Horizontal} = 1280.0{,}0015° \qquad (3.3)$$

$$FOV_{Vertical} = 960.0{,}0015° \qquad (3.4)$$
$$\approx 1{,}4°$$

TABLE II
CALCULATED DESIGN PARAMETERS OF OPTICAL SYSTEM

| Parameter | Calculated Data |
|---|---|
| Calculated FOV | 1,4º (V) x1,9º (H) |
| Calculated IFOV | ~0,0015º |
| Calculated Depth of Field | ~ 8,2 cm |
| Calculated Depth of Focus | ~ 1,7 mm |

The optical design is optimized using the design values listed in Table 1 and we manage to get a good spatial resolution for the source wavelength at 860 nm. The general layout of the designed system is given Fig.3. The spot diagram and the MTF of the designed optical system in in-focus conditions are given in Fig.4 and Fig.5, respectively.

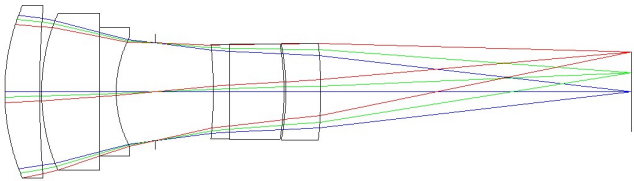

**Fig. 3.** Layout of the optical design

In-focus condition spot diagrams for different object angles are given in Fig.4. and in-focus condition MTF graph is given in Fig.5. In Fig.4. it can be seen that the spot sizes are smaller than the airy disk which means that optimized design is diffraction limited and high contrasted clear image is expected for in-focus conditions. Also, the OTF value for 116 cycles/mm is more than 0,55 in Fig.5., so a sharp image is expected for in-focus conditions.

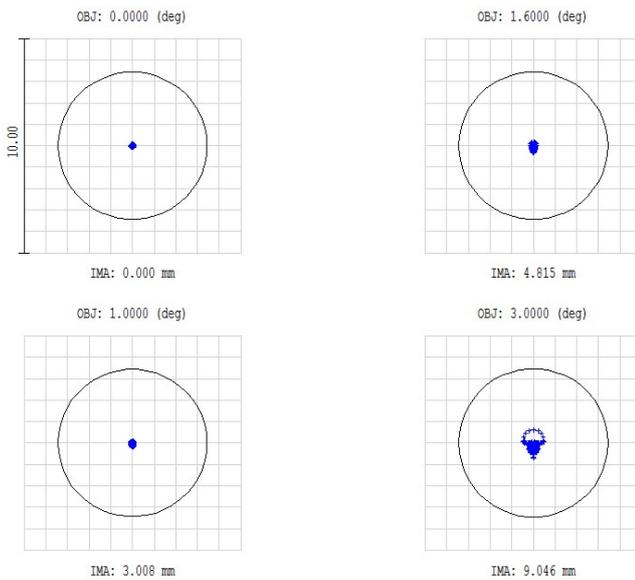

**Fig. 3.** Spot diagrams for the in-focus condition

For different depths or heights on the target plane, it is expected to observe slight focus change at the image plane. This de-focusing on the image plane will cause an intensity change on the detector. By using this, the depths or heights on the target plane can be measured by the measurement of the intensity change on the detector pixel. In Fig.6., the result of defocusing can be seen on the OTF. As a result, a depth change on the target plane will cause approximately 0,08 mm focus shift in the image plane and that will decrease the OTF value from 0,59 to approximately 0,05 as seen in Fig.6.

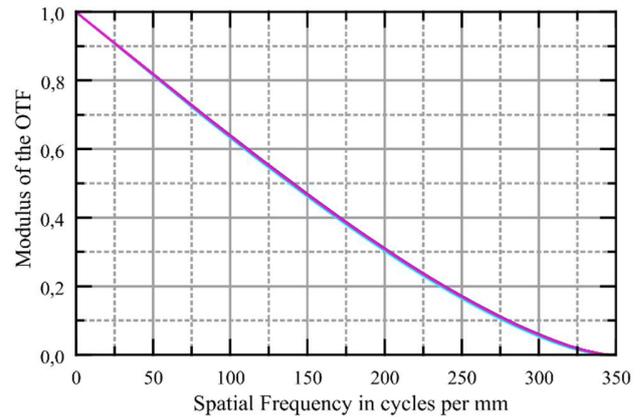

**Fig. 5.** MTF graph of the designed optical system at in-focus condition for Diffraction Limit, 0, 1.6 and 3.4 degree fields

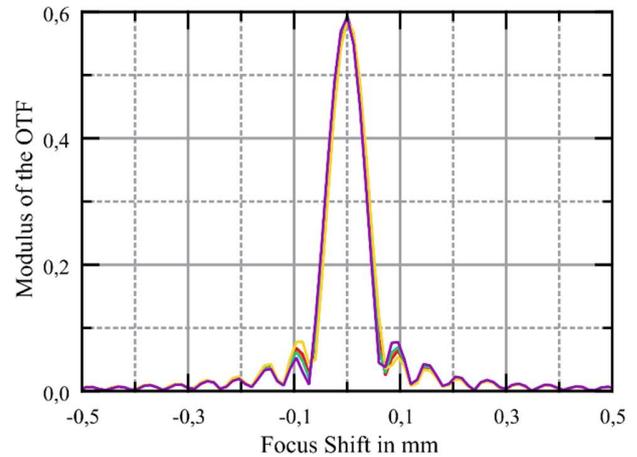

**Fig. 6.** OTF Modulus change with a shift in focus at the image plane

The change in the changing spot sizes on the image plane regarding de-focusing on the image plane is also given in Fig.7. The size of the small squares in the Fig.7. is 4 μm x 4μm which is very similar to the pixel size of the selected FPA. It can be easily seen that the light is focused on a single pixel in focus conditions where the light is spread over approximately 15 pixels in 50 μm de-focusing condition and the light is spread over approximately 52 pixels in 100 μm de-focusing condition. To observe how the target distance (depth on the target plane) is affecting the blurring on the image plane, the intensity changes on each pixel (encircled energy) due to object depth on the target plane is investigated with high resolution.

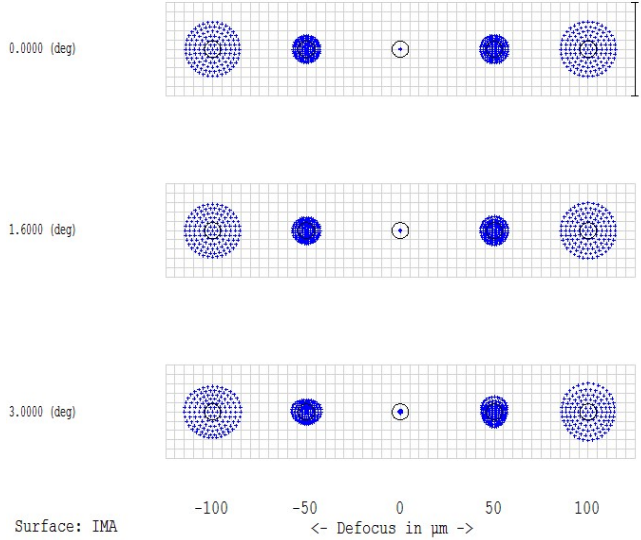

**Fig. 7.** Spot Size change regarding the defocusing

In Fig.8., it can be seen that ~%40 intensity loss occurs when target depth is changed from 40 mm to 60 mm (20 mm) on the target surface. In other words, the peak intensity condition for a point at 40 mm depth on the target plane is shifted ~34 μm according to Fig.6. So the reverse application will also be applicable for measuring the depth data such as shifting the image plane on the optical axis with the help of a high-resolution positioner. With this proposed method, measurement of the pixel intensity at different image plane locations will be collected for the estimation of the depth information on the target plane. Although, the reachable depth resolution is dependent on detector noise and false signal rates, at the optimum measurement point 1 mm depth change on the object plane is creating approximately %2 intensity change on the detector pixel which can be easily measurable with 8-bit detectors.

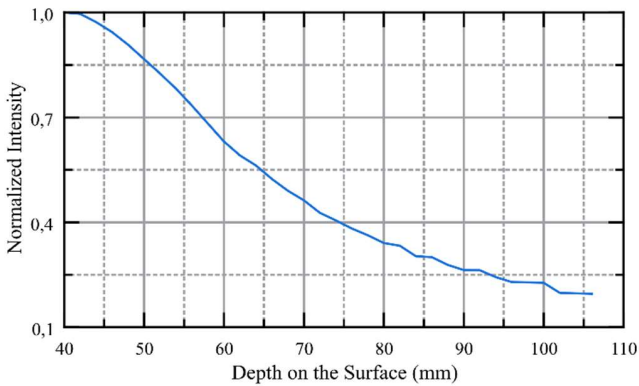

**Fig. 8.** Received Normalized Pixel Intensity change regarding focus shift on the target plane

## IV. THE DESIGN OF THE MULTI-DEPTH LIDAR SCHEME

For the proof-of-concept design, a basic optical system is designed and spot size changes at different image planes are worked for different target distances (depths). The basic design layout of the system incorporates three different pin-hole sensors on different image planes for resolving the different images that correspond to different target depths. The design of the multi-depth LIDAR scheme is shown in Fig.9.

In this study, separated three-mirror coated pin-hole sensor systems are used for measuring depth information with the images at different image plane locations. The mirror coating at the surface of the pin-holes provides the reflection of the non-focused beam rays to the following sensor plane. The physical placement requirements of the sensors have been taken into account and then necessary displacements between the sensor planes are provided for the placement of the sensors.

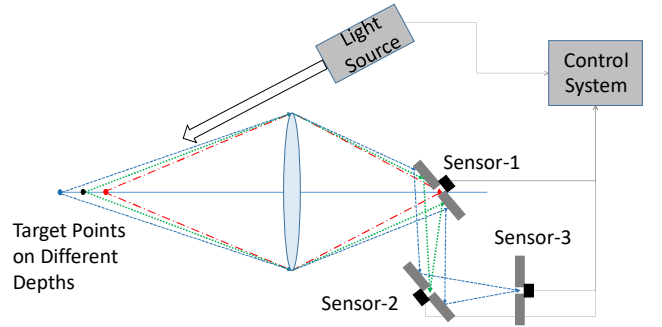

**Fig. 9.** Schematic diagram of the multi-depth LIDAR scheme with single-pixel detectors that are separated at the image plane. The reflective pinholes redirect the beam to the following detector and provide simultaneous multifocal imaging.

The first sensor is used for the time of flight measurement for the reference target distance at the first measurement step where all three sensors are used for imaging different depths on the target. Here, we aim to observe 2 mm depth resolution. An optical system is designed in OpticStudio and sensors are modeled on the image plane locations that are calculated for 2mm pre-determined depth resolution. The distance between the pin-hole sensors ΔZ is approximately M2Δz [5], where Δz is depth resolution and M is the magnification. In this proof of concept design, the distances are found as approximately 4,5 mm between each sensor plane which enable real-life implementation of our design, as shown in Fig.10.

The analysis of the depth information can be derived by the images at the different image planes; for predetermined 0 mm, 2 mm, and 4 mm heights on the target surface. For a target at 0 mm height in the object plane, it is observed that the collected light focused perfectly on the first sensor, S1. The signal on S1 after the pinhole provides a focal image. The remaining signal is reflected and directed to the second detector.

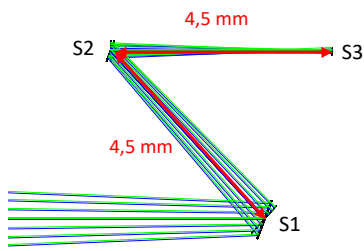

**Fig. 10.** Pin-Hole locations on Image Plane and beam sizes on pin-holes for an object on target plane (height= 4 mm)

For the target at 2 mm height in the object plane, the collected rays are reflected by the pinhole reflector that is in front of the first sensor. Similar to the first detector a focal signal on the second sensor plane will be observed. The remaining signal is again reflected by the reflective pinhole in front of the second detector and directed to the third detector.

Lastly, for the target at 4 mm height in the object plane, the reflected rays from the first and second sensor are focused on the third sensor plane. This result will gather low intensity on the S1 and S2 where high intensity on S3.

## V. RESULTS

The optical path and spot size results for different target distances are given in Fig.11.

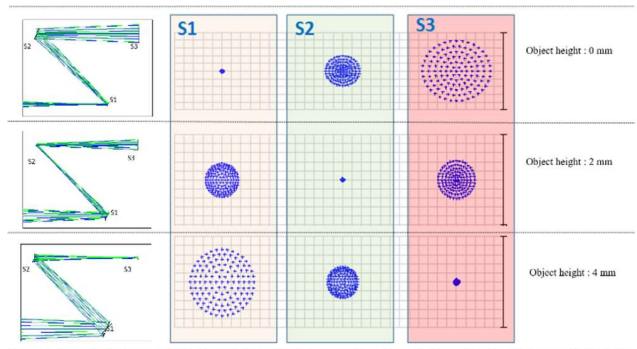

**Fig. 11.** Optical path and spot sizes on the s1, s2, and s3 sensors for different target distances

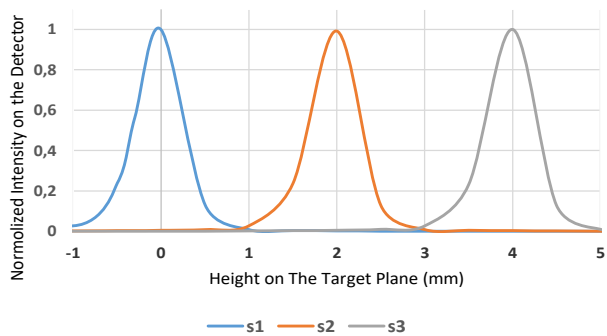

**Fig. 12.** The Normalized Encircled Energy change on Pinholes of Sensor-1 (s1), Sensor-2 (s2) and Sensor-3 (s3) via target point heights on the target plane

The encircled energy is also calculated at each sensor plane for different target distances (depths) and 6,8 μm pinhole sizes. The results for encircled energies are given in Fig.12. In encircled energy graph it can be seen that our concept is perfectly functional. The maximum encircled energy (intensity) is changing on each sensor plane regarding different target depths on the object plane. In Fig.12. it can be seen that different depths on the target plane can be distinguished easily with 2 mm resolution. The number of detectors can be increased further to increase the depth resolution.

Fig.11. shows the ray-tracing results and spot size results in each sensor plane for different target depths. For imaging an object at 0 mm height on the target plane, the light perfectly focused on S1 so, a clear image will be constructed in the S1 plane where there will be no or low-intensity images on the S2 and S3 planes as given in Fig.12. In other words, the image constructed by the S1 sensor data will give an image of the objects on the object plane. For an object at 2 mm height on the target plane, the light is de-focused on S1 and a bigger portion of the light is reflected by the mirror surface of the S1 pinhole. The reflected light is perfectly focused on S2 so, a clear image will be constructed in the S2 plane where there will be no or low-intensity images on the S1 and S3 planes as given in Fig.12. Then the image constructed by the S2 sensor data will give a sharp image of the objects at 2 mm height on the object plane where some very low-intensity blur images of objects at 4mm and 0 mm height on the target plane.

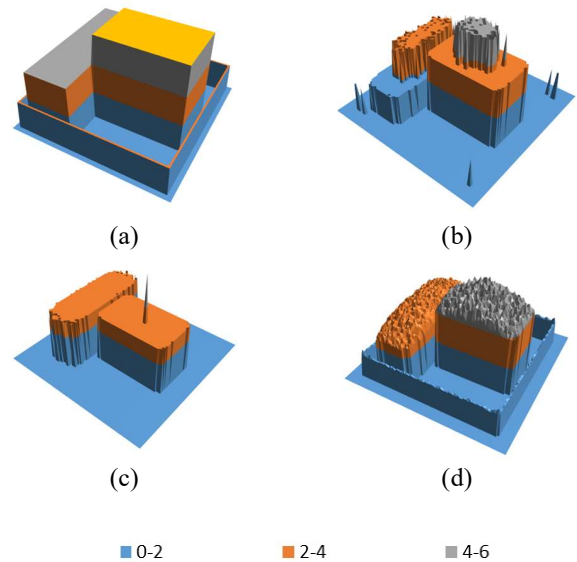

**Fig. 13.** Test sample, Simulated Lidar images and Multi-Depth Lidar system image. Thickness of each different colored layer is 2mm. **(a)** Test sample **(b)** Simulated Lidar image for 6 picoseconds Lidar Pulse Rise Time **(c)** Simulated Lidar images for 10 picosecond Lidar Pulse Rise Time **(d)** Simulated 3D image results of the Multi-Depth Lidar system for 4,5 mm spacing between the sensors.

For the further analysis, a test sample has been created with 2 mm depth differences between the surfaces as given in Fig.13. (a). Lidar images of the test sample for 10 and 6 picoseconds pulse rise times have been simulated for the best case scenario with the assumption that detector and electronic noises are negligible, and the results are given in Fig.13. (b) and (c). Finally, the images on each sensor planes are simulated via optical design program and sensor images are combined to create the 3D image of the test sample. The simulated 3D image results of the Multi-Depth Lidar system is given in Fig.13. (d). The multi-depth image can be reconstructed using pulse duration exceeding ns duration which is far beyond the resolution limits that can be reached with TOF measurements.

## VI. Conclusion

We show that mm depth resolution 3D imaging can be achieved by using a hybrid method that uses ToF measurement and depth measurement from focal shifting measurements without complicated picosecond pulsed laser systems or high bandwidth sensors. In this study, a proof of concept design has been shown and the design can be optimized for different measurement distances and different depth resolutions. Our design paves the way for a simpler and cost-effective method for 3D imaging systems that use single pixel detectors at single photon sensitivity level.

With this work, it is shown that limited depth of field can be used to generate a measurement window for the pre-defined distance measurement intervals. However, the depth resolution is limited by the de-magnification of the optical system for the longer ranges. The well-known AM modulated LIDAR techniques can be combined with the depth of field technique to get higher depth resolution by avoiding the ambiguity problem of AM modulation by the depth of focus effect. This developed technique can also be combined with scanner systems or micromirror scanner systems [10] for increasing the spatial resolution.

## Acknowledgment

We thank Nanomagnetics Instruments for the Zemax time they provided for our research.

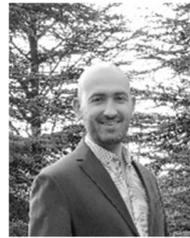

**Emre Yüce,** obtained his bachelor's degree from METU Physics department, Ankara/Turkey. He worked on "Electro-Optical Modulation of Silicon Microspheres" during his master's degree at Koç University, İstanbul/Turkey. He started his PhD degree at AMOLF institute in Amsterdam and obtained his PhD degree from University of Twente in Enschede/The Netherlands on "Ultimate-fast All-optical Switching of Microcavities". Between 2013-2106 he worked at MESA+ Institute for Nanotechnology as a post-doctoral researcher. He is currently working as an Assoc. Prof. Dr. at Middle East Technical University, Ankara/Turkey. His main research is focused on programmable photonics; optical computation, optical computation and imaging.

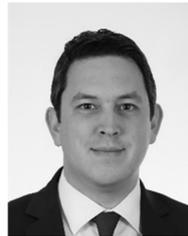

**Koray Ürkmen**, received his bachelor's degree from Izmir Institute of Technology Physics Department, Izmir/Turkey, in 2003 and the M.S. degree from Bilkent University Physics Department, Ankara/Turkey, in 2005. He is currently pursuing the Ph.D. degree in Middle East Technical University Micro and Nantoechnoly Department, Ankara/Turkey. His research interest includes the visible, IR and NIR optical systems, high resolution Lidar systems. Mr. Ürkmen is also working in industry since 2003 respectively on Probe Microscopy, Optical System Design, System Engineering and Business Development.